\documentclass[preprint,showpacs,amsfonts,amssymb,amsmath,12p]{revtex4}
\usepackage{times}
\usepackage{graphicx}

\begin{document}

\title{Casimir Friction II: Casimir effect and drag in zero-temperature superfluids}

\author{ David C. Roberts$^1$ (dcr@lanl.gov) and Yves Pomeau$^{1,2,3}$ (pomeau@math.arizona.edu)}  

\affiliation{$^1$ Theoretical Division and Center for Nonlinear Studies, Los Alamos National Laboratory, Los Alamos, NM \\ $^2$Department of Mathematics, University of Arizona, Tucson, AZ \\ 
$^3$ Laboratoire Physique statistique, Ecole normal Superierure, Paris, France} 

\maketitle

As discussed in the accompanying article \cite{cf1}, Casimir first predicted that zero-point quantum fluctuations of an electromagnetic (EM) field give rise to an attractive force
 between two closely spaced perfect conductors \cite{casimir}.  In this
 idealized situation, a macroscopic physical manifestation, i.e. a force, arises
 purely from the EM quantum fluctuations.
  Similarly, one would expect an analogous force to arise from quantum
 fluctuations in a quasiparticle vacuum in a zero-temperature dilute
 Bose-Einstein condensate (BEC). 

In this article, an argument is presented for the existence of a Casimir-like drag force on a stationary object in a slow-moving BEC due to the scattering of quantum fluctuations. This force would be the dominant effect at speeds well below that determined by the Landau criterion \cite{landau} (the derivation of which ignores the possibility of scattered quantum fluctuations).   No direct EM analogy can be drawn for this situation because of the Lorentz invariance of the EM vacuum, i.e. no absolute rest frame (where the relative motion of an object can be measured) exists for an EM vacuum. 
Before exploring the Casimir-like drag force, the connection between the familiar EM Casimir force and the Casimir-like force in a static BEC situation shall first be made explicit.  

The following estimate of a Casimir-like pressure $p_C$ in a weakly interacting BEC will serve to motivate the existence of a force in a BEC vacuum due to quantum fluctuations.  Consider two infinitely thin and infinitely repulsive plates (acting in an analogous role to perfect conductors in an EM field) spaced a distance $b$ apart in the $z$-direction, \footnote[10]{In principle, laser fields could play the role of the conceptual plates and the force would manifest itself as a transfer of momentum to the photons.} immersed in a three-dimensional condensate characterized by an interparticle contact pseudopotential, $g
 \delta^{(3)}(r)$, where $g$ is determined by the 2-particle positive scattering length $a$ and the mass $m$ of the atoms, i.e. $g=4 \pi \hbar^2 a/m$.  Let us assume the condensate is dilute such that $\sqrt{\rho_0 a^3} \ll 1$ where $\rho_0$ is the condensate density.  In the homogeneous dilute condensate  
(without confining walls), the ground state energy can be written \cite{Lee,Bog} as  
$E_g(k)=\rho_0^2 g V/2+ 1/2 \sum_{(\bf{k} \ne 0)} [E_B(k)-E_0(k)-\rho_0 g]$  
\footnote[11]{There is an additional term arising from a pseudopotential  
argument \cite{Lee} that allows for this energy to converge.  However, for simplicity, an energy cutoff is imposed and the pseudopotential term in the following discussion is left out as it does not play a  
significant role.} where $k=|\bf{k}|$, $E_0(k)=\hbar^2 k^2/(2m)$,  $ E_B(k)=[(E_0(k))^2+2 \rho_0 g E_0(k)]^{1/2}$ is the Bogoliubov  
dispersion relation, $\rho_0$ is the condensate density, and $V$ is the volume of the condensate.  The  
pressure generated by the leading order interaction term is given by $p_0=g \rho_0^2/2$.  Subsequent terms describe the contribution from quantum fluctuations, or quantum depletion, to  
the ground state energy, which gives rise to a Casimir-like force.
Assuming the distance between the confining
 walls to be much greater than the healing length, i.e. $b \gg k_0^{-1}=(8 \pi \rho_0
 a)^{-1/2}$ \footnote[12]{Note that, in the opposite limit when $b \ll k_0^{-1}$, the gas acts as a nearly ideal gas in which the fluctuations would have negligible effect.}, Casimir's original calculation can be followed to determine
 the change in energy per area between the homogeneous
 case, which displays a continuous momentum spectrum, and the case with confining walls, which has a discrete  
momentum spectrum along $z$.  The derivative with respect to the distance yields the Casimir-like pressure
\begin{equation}
\label{cas}
p_C=-\frac{d}{d b} \left(  
\frac{1}{2}\chi(0)+\sum_{n=1}^{\infty}\chi(n)-\int_0^\infty dn \chi(n)\right), 
\end{equation}
where $\chi(n)=1/4 \pi \int_0^{\Lambda_{k_r}} k_r d k_r[E_B(n)-E_0(n)  
-\rho_0 g]$, $k^2(n)=k_r^2+(n \pi/b)^2$, and $\Lambda_{k_r}$ is the ultraviolet cutoff where the simple assumption of a delta-function contact pseudopotential breaks down.   
Using the Euler-MacLaurin formula and taking $\Lambda_k \rightarrow \infty$ gives  
the Casimir-like pressure in a dilute condensate to leading order as
\begin{equation}
p_C \approx -\frac{d}{d b} \left( \frac{1}{30 \cdot 4!}\chi'''(0)\right)  
\approx - \frac{\pi^2}{480}\frac{\hbar c_s}{b^4},
\end{equation}
where the derivative is with respect to $n$ and the speed of sound in this system is given by $c_s=\sqrt{\rho_0 g/m}$.  

The connection with the Casimir force is clear: this expression for the Casimir-like pressure is exactly one half of the original Casimir pressure arising from the zero-point fluctuations of the EM vacuum, with the speed of sound replacing the speed of light.  This connection can be understood in light of the observation that the lower wavenumber modes dominate the Casimir-like pressure (the higher modes cancel out as the discrete momentum spectrum becomes more like a continuous spectrum).  At long wavelengths the ground state energy is approximately $E_g(k)=\rho_0^2 g V/2+ 1/2 \sum_{(\bf{k} \ne 0)} [\hbar c_s k-\rho_0 g]$, which, aside from an unimportant constant factor and the substitution of the speed of light with $c_s$, is the same harmonic oscillator energy spectrum that describes the EM vacuum.  The factor of one half between the BEC case and the EM case arises as there are two transverse polarization states for each EM mode whereas there is only one polarization state for the phonon modes in a BEC.  Although for a dilute condensate this force is $10^6$ times smaller than the Casimir force in an EM vacuum and finite-size effects present in experiments have not been included, this force could in principle be measured, providing direct evidence of the quantum fluctuations in these systems.  

The next step is to show that a drag force also arises from these fluctuations in a slow-moving BEC flow around a stationary object.  Since this force arises from quantum fluctuations, one needs to go beyond the usual mean field picture and treat the fluctuations as scattered waves.   An outline of this calculation for a general object is presented below (for more details see \cite{roberts,roberts2}).

In general, a force on a stationary object described by a potential $V(r)$ ($r$ represents a vector quantity throughout this paper) can be written in second quantized notation at zero temperature as
\begin{equation}
{\vec F}=-\int d^3r \langle \hat \psi^\dag (r)[ {\vec \nabla} V(r)]\hat \psi (r) \rangle_{T=0},
\end{equation}
where $\hat \psi(r)$ and $\hat \psi^\dag(r)$ are field operators that describe the weakly interacting BEC flow and obey the standard boson commutation relations (the expectation value at $T=0$ is defined below).
 Landau, based on his phenomenological theory that did not explicitly consider the quantum fluctuations, showed the moving ground state to decay by the emission of quasiparticles only above a critical speed; below that critical speed the flow would be dissipationless \cite{landau}.  Consequently, according to Landau's theory a stationary object in a superfluid flowing below this critical velocity would remain metastable and not experience any force.  

Landau's argument, however, only directly applies to a uniform condensate (containing infinitesimally small test particles that do not perturb the condensate flow) whose state is derived from the immobile ground state by a Galilean transformation.  The critical velocity in a non-uniform system is more complicated as the quantum state is not derived by a simple uniform Galilean boost.  At zero temperature it can generally be assumed that the condensate, whether uniform or non-uniform, is well described by a macroscopic classical field, i.e. $\hat \psi (r)= \Psi^{(0)}(r)$.
The force on a stationary object in a moving flow is then simply given by
\begin{equation}
{\vec F}_{GPE}=-\int d^3r |\hat \Psi^{(0)}(r)|^2 {\vec \nabla} V(r) 
\end{equation}
and the behavior of this field in a time-independent situation is given by the Gross-Pitaevskii equation (GPE) 
\begin{equation}
\label{GP} (\hat T+V(r)-\mu) \Psi^{(0)}(r)+g | \Psi^{(0)}(r) |^2
\Psi^{(0)}(r)=0,
\end{equation}
where $\mu$ is the chemical potential, $\hat T \equiv - \frac{\hbar^2 \nabla^2}{2 m}+i \hbar c \frac{\partial}{\partial x}+\frac{1}{2} mc^2$ and $c$ is the speed of the flow far from the object.  In the non-uniform case, it is the maximum local velocity that should be compared to Landau's critical velocity.  Using the GPE, one can show that $F_{GPE}$ is zero only for a flow velocity at infinity that is associated with a maximum local flow velocity that does not exceed Landau's critical velocity.  Although the critical velocity, as measured at infinity, is modified in the non-uniform case, the principle of Landau's argument still applies.  However, if quantum fluctuations | beyond the scope of the GPE | are included then even such a modified critical velocity defining a transition to frictionless flow no longer necessarily applies to this situation.  Hereafter, to avoid confusion in the discussion below, no explicit distinction will be made between the modified critical velocity in the non-uniform case and Landau's critical velocity in the homogeneous case.

Below, the calculation for the force due to the quantum fluctuations at $T=0$ is described.  The assumption is made that the superfluid is well approximated by a weakly interacting  Bose-Einstein condensate.  The Bose field operator that describes the flow is approximated as a large macroscopic classical field (as above) plus a small quantum fluctuation operator (the small parameter in this case is given by the diluteness parameter $\sqrt{\rho_0 a^3}$ \cite{fetter}), i.e.
\begin{equation}
\hat \psi (r)= \Psi^{(1)}(r) +\hat \phi(r),
\end{equation}
where $\Psi^{(1)}$ is the macroscopic condensate that has been modified by the quantum fluctuations. 
The force acting on a stationary potential can then be written as
\begin{equation}
{\vec F}=-\int d^3r (| \Psi^{(1)}(r)|^2+ \langle \hat \phi^\dag (r) \hat \phi (r) \rangle_{T=0}){\vec \nabla} V(r).
\end{equation}

Next, $\hat \phi$ \cite{Bog} is expanded in terms
of quasiparticle operators $\hat \alpha_{k}$ and $\hat \alpha^\dag_{k}$ that, by
definition, obey the standard boson commutation relations $[\hat
\alpha_k,\hat \alpha^\dag_{k'}]=\delta_{k,k'}$ where $\delta_{k,k'}$ is the Kronecker delta function.
The quasiparticle
operators are weighted by the quasiparticle amplitudes
$u_{k}(r)$ and $v_{k}(r)$, i.e.
\begin{equation}
\label{qp} \hat \phi(r)=\sum_{(\bf{k} \ne 0)} \left( u_{k}(r) \hat
\alpha_{k} -v^*_{k}(r) \hat \alpha_{k}^\dag \right), 
\end{equation}
where the sum is taken over all excited states and excludes the condensate mode and $*$ denotes the complex conjugation.   The assumption is made that the quasiparticle amplitudes 
satisfy the two sets of differential equations \cite{fetter}, also known as the Bogoliubov equations,
\begin{equation}
\label{BdG1} \hat {\cal L} u_{k}(r) -g(\Psi^{(0)})^2 v_{k}(r) =
E_k u_{ k}(r)
\end{equation}
\begin{equation}
\label{BdG2} \hat {\cal L}^*  v_{k}(r)- g(\Psi^{(0)*})^2 u_{k}(r)
= -E_k v_{ k}(r),
\end{equation}
where $*$ denotes the complex conjugation, $E_k$ is the eigenvalue associated with momentum state $\bf{k}$,
and $\hat {\cal L} = \hat T + V(r) - \mu +2 g|\Psi^{(0)}|^2$. The quasiparticle operators would then diagonalize the weakly interacting Hamiltonian.  In other words, weakly interacting particles are transformed into non-interacting quasiparticles.

The behavior of the condensate modified by the quantum fluctuations at $T=0$ is given by the generalized GPE \cite{Castin}
\begin{equation}
 (\hat T+V(r)-\mu) \Psi^{(1)}(r)+g | \Psi^{(1)}(r) |^2 \Psi^{(1)}(r)
+ \sum_{(\bf{k} \ne 0)} 2 g |v_k(r)|^2\Psi^{(1)}(r)-g \sum_{(\bf{k} \ne 0)}  u_k(r)v^*_k(r)\Psi^{(1)}(r)-f(r)\Psi^{(1)}(r)=0.
\end{equation}
The term proportional to $\sum_{(\bf{k} \ne 0)} u_k(r)v^*_k(r)$ is ultraviolet divergent because of the contact potential approximation and must be renormalized. The last term, $f(r) \Psi^{(1)}(r)$, ensures the orthogonality between the excited modes and the condensate and is given by $f(r)=\sum_{(\bf{k} \ne 0)} c_k v^*_k(r),$
where $c_k=g \int d^3r  | \Psi^{(1)}(r) |^2(\Psi^{(1)*}(r) u_k(r)+ \Psi^{(1)}(r) v_k(r))$.

Since the fluctuations are assumed to be small, the fluctuation part of the condensate can be separated from the macroscopic part described by the GPE, i.e. $\Psi^{(1)}(r)=\Psi^{(0)}(r) +\sum_{(\bf{k} \ne 0)} \delta \Psi_k(r)$, where $\delta \Psi_k(r)$ is associated with a given mode $k$.  Therefore, since it is known that $F_{GPE}=0$ at speeds lower than the critical velocity, at these low speeds the force (at the dominant order) on a stationary object can be written as  
\begin{equation}
{\vec F}_{fluc}=-\sum_{(\bf{k} \ne 0)} \int d^3r (\Psi^{(0)}(r) \delta \Psi_k(r)^*+c.c.+ |v_k(r)|^2){\vec \nabla} V(r),
\end{equation}
which is proportional to the diluteness parameter $\sqrt{\rho_0 a^3}$.

To complete the calculation of the Casimir-like force, first the coupled Bogoliubov equations (eqns \ref{BdG1}, \ref{BdG2}) must be solved for the  quasiparticle amplitudes, then the generalized GPE must be solved for $\delta \Psi_k(r)$, and finally all excited states must be integrated over.  The Bogoliubov equations can be treated as a scattering problem in which the only incoming waves considered are the incident homogeneous fluctuations (incoming scattered waves are excluded); exponentially growing scattered waves are excluded.   Temperature is not well defined in this scattering problem, so instead one can impose in terms of measurable quantities (specifically the spatio-temporal correlation functions) a condition far from the potential that determines the quantum state in this problem (see \cite{roberts} for details) such that the state is annihilated by $\hat \alpha$.   Therefore, it remains convenient to denote that state by `$T=0$'.  The expectation values  written in terms of quasiparticle amplitudes thus  take the form $\langle \hat \phi^\dag (r) \hat \phi (r) \rangle_{T=0} = \sum_{(\bf{k} \ne 0)} |v_k(r)|^2$, for example.

The calculation in the Born approximation is described in \cite{roberts} for a specific potential that varies only in the flow direction, and in \cite{roberts2} for a 3-D weak point impurity, both in an infinitely extended medium.  In the latter case, the force is shown to be linearly proportional to the flow velocity.  A force is found to exist at all flow velocities in both cases.

${\vec F}_{fluc}$, a direct physical manifestation of the zero-point fluctuations, dominates at velocities below Landau's critical velocity because, due to the Landau criterion, there is no contribution described by the GPE (eq. \ref{GP}) to the overall force.  ${\vec F}_{fluc}$ comes not from a change in the number of quasiparticles, but from a change in the associated amplitudes resulting from the scattering condition necessary to solve the Bogoliubov equations. Assuming this force to be dissipative \footnote[13]{While one would expect this force to be dissipative (this can be proven by computing the time-dependence of flow energy in the lab frame for example), the existence of this force is a necessary but not sufficient condition for the system as a whole to be dissipative.}, energy is transferred by the scattering of fluctuations (analogous to wave drag in classical fluid mechanics) rather than by the creation of quasiparticles at higher flow velocities or the scattering of quasiparticles at non-zero temperature.

Although the existence of a drag force on a localized object in an infinitely extended BEC flow seems to imply that persistent currents cannot exist, this effect {\it can} be consistent with the observed persistent currents, where superfluids placed in a closed loop are able to flow practically endlessly without any detectable friction.  Consider a closed loop of flowing dilute condensate at zero temperature contained by perfectly smooth walls (we consider rough walls below) with an impurity fixed in the flow.   From the discussion above, one would expect (ignoring typical transient effects, unrelated to quantum fluctuations, that arise from the act of placing the impurity in the flow) that for a short time interval from the moment the impurity is introduced in the flow, this system would act as if the medium were infinitely extended, i.e. the impurity would feel a force and the flow would dissipate energy.  However, the scattering fluctuations experience the finitude of the system at time scales on the order of the characteristic length of the system divided by the speed of sound.  The scattering fluctuations then re-interact with the impurity, eventually cancelling out any drag force.   A certain amount of scattered fluctuations must be present before this force can cancel out.  If one includes higher order terms in the Hamiltonian that describe the quasiparticle interactions (i.e. the Beliaev and Landau damping terms), one can assume the system will relax to a local thermal equilibrium.  In other words, the scattered fluctuations will relax to a small amount of normal gas.  Therefore, one would expect there to be two relevant observables | a new time scale for the finitude of the system to be felt and a small amount of normal gas that must be produced to counteract the force on the fixed impurity far from the object.

Finally, one expects the scattering of quantum fluctuations off a rough surface to have dramatic effects on the superfluid boundary conditions.   As discussed in \cite{bc}, in addition to the total normal mass flux condition for a solid wall, $\vec{n} \cdot (\vec{j}_n + \vec{j}_s) = 0$ where $n$ is the normal vector to the surface and $j_n$ and $j_s$ are the mass fluxes of the normal and superfluid, the scattering of quantum fluctuations off a rough surface leads to the following non-equilibrium hydrodyamic boundary condition
\begin{equation}
\vec{n} \cdot (\vec{j}_n - \vec{j}_s)= \alpha (T_b-T) +\beta v_s^2,
\end{equation} 
where $T_b$ is the temperature at the boundary, and $T$ is the temperature of the fluid nearby.  This boundary condition holds to leading order and has important predictions, such as a new critical speed and that there should be a temperature difference between a moving superfluid and the boundary, which implies that there should always be at least a small amount of normal fluid detectable in immersed torsion pendulum experiments.

\vspace{1 in}
{\large Mathematical Subject Classifications}
\begin{itemize}
\item 82D50 Superfluids
\item 81U99 Scattering Theory
\item 81V70 Many-body theory; quantum Hall effect
\end{itemize}

\end{document}